\documentclass{nature}

\usepackage{caption}
\usepackage{mathtools}

\usepackage{enumitem}
\usepackage{xr-hyper}
\usepackage{hyperref}

\usepackage{lineno}
\usepackage{siunitx}
\usepackage{amsmath}
\usepackage{amssymb}
\usepackage{graphicx}
\usepackage{placeins}

\makeatletter
\makeatletter

\usepackage[utf8]{inputenc}
\makeatletter
\let\saved@includegraphics\includegraphics
\AtBeginDocument{\let\includegraphics\saved@includegraphics}
\renewenvironment*{figure}{\@float{figure}}{\end@float}
\makeatother
\title{Planarized THz quantum cascade lasers for broadband coherent photonics}
\author{Urban Senica$^{1,*}$, Andres Forrer$^{1}$, Tudor Olariu$^{1}$, Paolo Micheletti$^{1}$, Sara Cibella$^{2}$, Guido Torrioli$^{2}$, Mattias Beck$^{1}$, J{\'e}r{\^o}me Faist$^{1}$, Giacomo Scalari$^{1,*}$}

\begin{document}

\maketitle
\begin{affiliations}
 \item Quantum Optoelectronics Group, Institute of Quantum Electronics, ETH Z{\"u}rich, 8093 Z{\"u}rich, Switzerland
 \item Istituto di Fotonica e Nanotecnologie, CNR, Via del Fosso del Cavaliere 100, 00133 Rome, Italy

\end{affiliations}

\begin{abstract}
Recently, there has been a growing interest in integrated THz photonics for various applications in communications, spectroscopy and sensing. We present a new integrated photonic platform based on active and passive elements integrated in a  double-metal, high confinement waveguide layout planarized with a low-loss polymer. An extended top metallization results in low waveguide losses and improved dispersion, thermal and RF properties, as it enables to decouple the design of THz and microwave cavities. Free-running on-chip quantum cascade laser combs spanning 800 GHz, harmonic states over 1.1 THz and RF-injected broadband incoherent states spanning over nearly 1.6 THz are observed. With a strong external RF drive, actively mode-locked pulses as short as 4.4 ps can be produced, as measured by SWIFTS. 
We demonstrate as well passive waveguides with low insertion loss, enabling the tuning of the laser cavity boundary conditions and the co-integration of active and passive components on the same THz photonic chip. 

\end{abstract}


\section*{Introduction}
 Integrated photonics \cite{RoadSilPhot_2016} makes extensive use of on-chip optical elements such as sources, splitters, modulators,  and high-confinement waveguides embedded in a planar platform to efficiently process and route optical signals. There is a growing interest in integrated Mid-IR and THz photonics for telecommunications and sensing \cite{Smit_IntegratedInP,SenguptaNatElec2018}. In the THz frequency range, a prominent candidate for source integration is the THz quantum cascade laser \cite{kohler_terahertz_2002}. Recent advances in the high temperature operation of these devices,  \cite{bosco_thermoelectrically_2019,khalatpour_high-power_2020} combined with  their frequency agility \cite{Curwen_NatPhot2019} and the possibility to operate as frequency combs \cite{burghoff_terahertz_2014,rosch_octave-spanning_2015} as well as very fast detectors \cite{Micheletti2021} make them extremely appealing as key building blocks for THz photonics. 
 
 Crucial features for laser integration \cite{RoelkensBowers_LPR2010_lasintegr} in more complex photonic systems \cite{Piccardo_Roadmap_2021} are the reduction of the electrical consumption, and consequently of the injected current, and the efficient coupling to low-loss passive waveguides.  Here, we propose a new platform for integrated THz photonics that allows signal propagation with passive elements and coherent source integration for applications such as broadband sensing \cite{ConsolinoHybrid2020} and coherent telecommunications \cite{Koenig_Koos_NatPhot2013}. In this first demonstration, we leverage the presence of a common metallic ground plane to demonstrate the integration of several active and passive THz photonic components onto the same semiconductor platform, allowing for efficient signal processing at THz and RF frequencies. 
 
 \section*{Planarized waveguide platform}
\par
\begin{figure}[h!]
\centering
\includegraphics[width=0.8\linewidth]{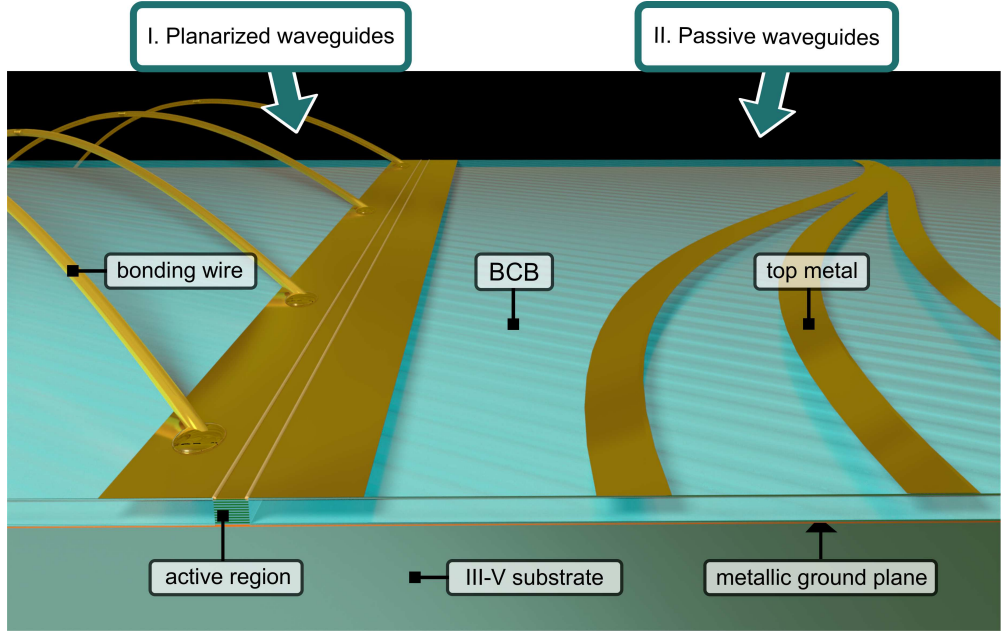}
\caption{Our new platform for broadband coherent THz photonics is based on planarized active and passive waveguides embedded in BCB, a low-loss polymer. \textbf{I.} The planarized active waveguide consists of a standard double metal waveguide encompassed in BCB and an extended top contact metallization. With bonding wires placed over the passive section, improved waveguide losses, RF and thermal properties are observed (see text for details). 
\textbf{II.} Passive waveguides can also be fabricated, consisting of metallic stripes on top of BCB, which provide confinement and guide the optical mode. These can be used to co-integrate active and passive elements on the same chip.
}
\label{fig:wg_schematic}
\end{figure}

The basic building block is a high-performance planarized double-metal waveguide with an extended top metallization, as shown in Fig. \ref{fig:wg_schematic} on the left. A similar kind of waveguide has already proven to be very effcient both for THz and microwave applications \cite{Maineult_APL_2010}. Following a standard double metal waveguide fabrication process \cite{williams_terahertz_2007} with dry-etched active region waveguides, a microelectronic-grade low-loss polymer Benzocyclobutene (BCB) is spin-coated and baked as the surrounding material (see Methods for details). The latter is widely used in microelectronics and has already been successfully employed in several THz applications \cite{BCB_TDS_Perret, Bosco:2016dk}. Specifically, we use Cyclotene
3022-57 (BCB), with a refractive index of 1.57 and relatively low losses (3 cm\textsuperscript{-1} = 1.3 dB/mm) at 3 THz \cite{bonzon2016phase}, making it an ideal planarization material.   
To ensure a flat and smooth profile, the BCB spin and bake step is repeated five times, and the top surface is subsequently etched with RIE to the same height as the active region waveguide. An extended top metallization with a typical width of \SI{300}{\micro\metre} is deposited over the active region and the BCB-covered area on the sides, which offers several advantages, as discussed below. 
From the active region standpoint, for these first demonstrations we used a strongly diagonal, low-threshold broadband GaAs/AlGaAs heterostructure, fully described in Ref. \cite{forrer_photon-driven_2020}.

\par 
We start by presenting results on the integration of on-chip THz frequency combs making use of planarized waveguides and optimized RF input and output coupling. In the context of THz frequency combs, we have shown that the control over transverse modes is essential in order to obtain a regular and flat-top comb spectrum \cite{bachmann_short_2016}. The introduction of side absorbers  mitigates lasing in higher-order transverse modes due to increased waveguide losses. A similar result can be achieved by reducing the transverse dimension of the laser ridge to \SI{60}{\micro\metre} and below. However, for conventional double metal waveguides, the width cannot be arbitrarily small since the waveguides are usually contacted by wire bonding on the top metallic cladding. This inherently limits the effective ridge width $w$ to the dimensions of the bonding wire patch, making reliable devices with ridges of \SI{50}{\micro\metre} or below challenging  to contact and prone to failures. Bonding directly on the active region can moreover introduce defects, increasing the waveguide losses and non-intentionally selecting specific modes, potentially compromising the long-term performance of the device. 
\par With our planarized platform, all of these issues can be solved. Placing the bonding wires on top of the extended top metallization over the passive, BCB-covered area, prevents formation of any defects or local hotspots on top of the active region, and enables the fabrication of very narrow waveguides, well below the bonding wire size. The narrow waveguide width can be employed as an efficient selection mechanism for the fundamental transversal lasing mode, and is also beneficial for heat dissipation and high temperature continuous wave (CW) operation. With reduced waveguide widths we enter the regime of “wire” lasers \cite{amanti-Wirelas_opex2010}, which have a very favourable figure of merit for their surface-to-volume ratio as it scales as the inverse of the width S/V$\propto$1/w. The extended contact eases also the heat extraction as in a radiator scheme. 


 \section*{Simulations}

\par
\begin{figure*}[h!]
\centering
\includegraphics[width=0.95\linewidth]{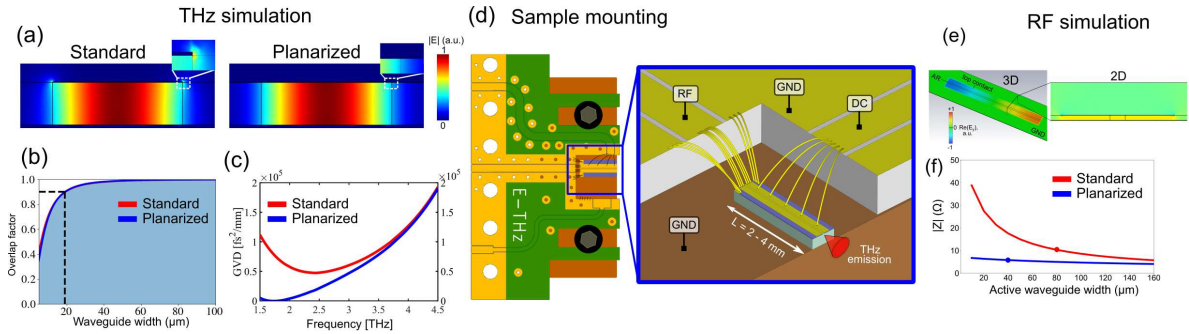}
\caption{\textbf{(a)} COMSOL 2D eigenmode simulation of a \SI{40}{\micro\metre} wide waveguide at 3 THz. While there are field spikes present in the standard waveguide at the corners, for planarized waveguides the field distribution is smooth also at the environment boundary. \textbf{(b)} Computed overlap factors with the active region are comparable for the standard and planarized waveguides. The dashed line indicates the width where the overlap factor drops below 90\%, for a frequency of 3 THz. \textbf{(c)} Computed chromatic dispersion for a waveguide width of \SI{40}{\micro\metre}. At low frequencies (3 THz and below), the planarized waveguide has a significantly lower GVD than a standard waveguide, where there are dispersive field spikes as the mode starts to leak out at lower frequencies. \textbf{(d)} Illustration of the sample mounting with a custom RF-optimized PCB. Several short signal wires are connected from a 50 $\Omega$ matched coplanar waveguide to one end of the waveguide, with short ground wires for minimal RF injection and readout losses (inset, top left). The DC laser bias is provided through a separate contact pad (inset, top right).  \textbf{(e)} 2D and 3D eigenmode numerical simulations of the RF field at 20 GHz show that the whole extended top metallization encompasses a cavity with the ground plane. \textbf{(f)} Computed impedance from 2D COMSOL simulations for a frequency of 20 GHz and a varying active waveguide width. Typical waveguide widths are marked with circles.}
\label{fig:simulations}
\end{figure*}

 
With the extended top metallization, the active region is sandwiched in a symmetric structure. As a consequence, the propagating optical mode does not feature any field spikes on the corners and at the edges  of the ridge, as is the case for a standard double metal waveguide, as illustrated in Fig. \ref{fig:simulations}(a). One major advantage of using double metal waveguides is the large overlap factor of the propagating mode, reaching nearly unity. It can, however, reduce with the waveguide width. In Fig. \ref{fig:simulations}(b), we show that the computed overlap factor remains above 90\% for ridge widths of above \SI{20}{\micro\metre} for both the standard and planarized waveguides at a frequency of 3 THz. Since we are interested in frequency combs, another important figure of merit is the group velocity dispersion of the waveguide. In Fig. \ref{fig:simulations}(c) we display results of COMSOL 2D eigenmode simulations, showing  that the dispersion of the planarized waveguides is significantly reduced with respect to the standard ones, especially at low ($<$3 THz) frequencies. This is again due to the absence of any field spikes which introduce dispersion, as the optical mode starts to leak to the sides in standard double metal waveguides. The computation includes waveguide and material (GaAs) dispersion for a \SI{40}{\micro\metre} wide ridge waveguide. 

\par
Heat dissipation properties are improved as well, as lateral heat transport takes place through the extended top metallization and the BCB polymer. COMSOL 2D thermal simulations show that for a \SI{40}{\micro\metre} wide waveguide, the maximum temperature inside the active region is reduced by around 7 K in a planarized waveguide at maximum bias conditions (11 V, 400 mA/cm\textsuperscript{2}) and a heat sink temperature of 100 K. A measurement study of the threshold current density as a function of increasing heat sink temperature \cite{faist_quantum_2013} and a comparison with data from our previous work \cite{forrer_rf_2020} also show a factor of 1.6 higher thermal conductivity of 130 W/Kcm\textsuperscript{2}. More detailed results of these simulation and characterization studies can be found in the Supplementary Material.
\par 
The lateral heat transport could be enhanced by functionalizing BCB with nanoparticles, as demonstrated in \cite{Xu2012}, without affecting the waveguide's optical losses. Moreover, as the planarized platform allows for the fabrication of even narrower waveguides, heating effects can be mitigated further.
Recent results also display the positive impact of a thinner active region in obtaining high temperature CW operation \cite{AIPAdvances_Curwen_2021}. This approach can be as well combined with our planarized geometry to push the high temperature operation even higher.

\par 
Next, we study the radio-frequency (RF) properties of the planarized waveguides. These are crucial both for the extraction and measurement of frequency comb beatnotes as well as for an efficient injection of RF signals which can affect and control laser operation. The relevant RF frequency range is close to the cavity repetition rate, defined as $f_{\mathrm{rep}}=\frac{c}{2\ n_\mathrm{g}\ L}$, where $c$ is the speed of light in vacuum, $n_\mathrm{g}$ the mode group index, and $L$ the cavity length. For typical waveguide lengths between 2-4 mm, the corresponding $f_{\mathrm{rep}}$ lies roughly between 10-20 GHz, although harmonic comb states can generate RF beatnotes well above 50 GHz. 
\par 
\par To fully capitalize on the improved RF properties, a dedicated RF PCB was developed and used for the laser mounting on copper submounts. As illustrated in Fig. \ref{fig:simulations}(d), the PCB features independent DC bias and RF readout/injection lines. The RF line is a straight 50 $\Omega$ matched coplanar waveguide which enables the placing of several short bonding wires at the backside of the laser waveguide, while short ground wires on each side close the loop to minimize RF losses. The positioning of all the RF signal wires on only one end of the waveguide is crucial in order to maximize the RF readout/injection efficiency, since the RF field produces a standing wave across the whole cavity with a node in the center and maxima at each end,\cite{piccardo_time-dependent_2018} as shown in the simulation in Fig. \ref{fig:simulations}(e). A separate dedicated, non-matched contact pad provides the DC bias to the device, with bonding wires distributed along the whole length of the laser waveguide for a homogeneous current injection.
\par 
2D and 3D eigenmode numerical simulations of the RF field of the planarized waveguide suggest that the whole extended top metallization encompasses a cavity with the ground plane, with the electromagnetic field oscillating across the whole patch, as shown in Fig. \ref{fig:simulations}(e) for a frequency of 20 GHz. Compared to standard waveguides, the RF properties of the planarized waveguides are modified in several aspects. First, due to the large modal overlap with BCB (low refractive index of 1.57), the effective index of the microwave mode $n_\mathrm{eff}^\mathrm{GHz}$ is reduced, which results in an increase of the fundamental microwave resonance frequency, detuning it from the THz mode spacing frequency $f_{\mathrm{rep}}$. Second, this also reduces the waveguide impedance at RF frequencies, as shown in Fig. \ref{fig:simulations}(f) for a frequency of 20 GHz and a varying width of the active waveguide. Here, we performed 2D COMSOL simulations of the waveguide cross section, where the refractive index of the active region was fixed to 3.6, while the metals were approximated as a perfect electric conductor (PEC). The lower impedance results in higher Q-factors of the microwave modes in the planarized waveguide, and in reduced radiative RF losses. Since the impedance of the planarized waveguide depends mostly on the width of the extended top contact, this allows for an independent design of the RF properties (e.g., by changing/modulating the width of the extended contact), while keeping the waveguide properties at THz frequencies unchanged. A more detailed analysis of RF properties with 3D numerical simulation results can be found in the Supplementary Material.

 \section*{Experimental results}

\par
\begin{figure}[h!]
\centering
\includegraphics[width=0.9\linewidth]{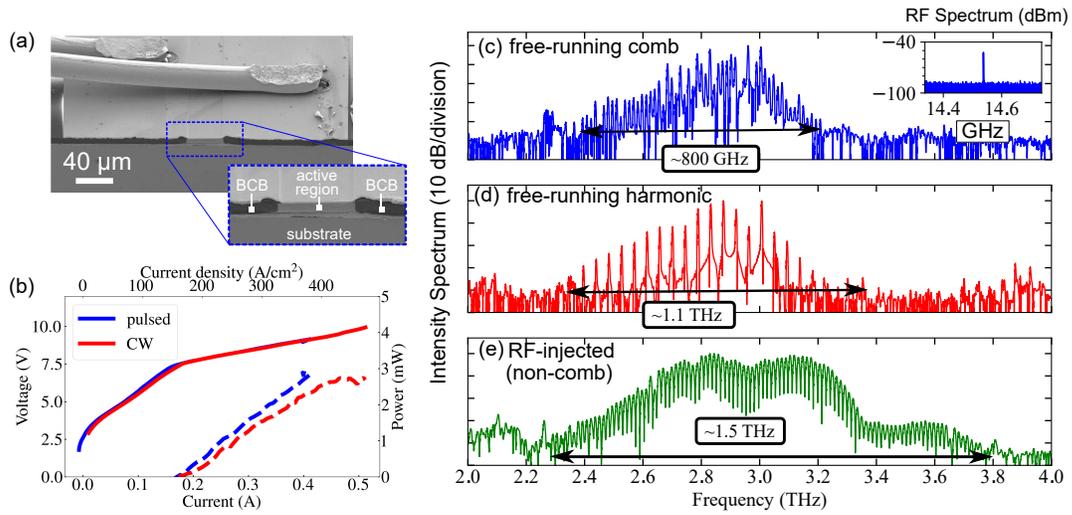}
\caption{ \textbf{(a)} SEM image of a planarized, \SI{40}{\micro\metre} wide ridge waveguide device, showing the cleaved front facet. The bonding wires are placed on the extended top metallization over the BCB-covered area. \textbf{(b)} LIV curves of a ridge laser sample measured in pulsed mode (500 ns pulses, 10\% duty cycle) and in CW at a heatsink temperature of 40 K. \textbf{(c)} A free-running frequency comb spanning around 800 GHz, with the measured single strong RF beatnote (-55 dBm) shown in the inset. \textbf{(d)} A broadband free-running third harmonic state, covering a bandwidth of 1.1 THz. \textbf{(e)} With strong RF injection (+32 dBm at source) close to the free-running mode spacing, the emission spectrum can be broadened, spanning around 1.5 THz.
}
\label{fig:RF_THz}
\end{figure}

 
We present now experimental measurements, first investigating the performance of simple ridge devices. These have a typical waveguide width of \SI{40}{\micro\metre}, narrow enough for fundamental transversal mode selection, and wide enough for a large overlap factor and low propagation losses. Similar as in the case of standard double metal waveguides, a Fabry-Pérot cavity is formed by mechanical cleaving, which forms atomically flat end facets. In Fig. \ref{fig:RF_THz}(a) we show an SEM image of the front part of the ridge waveguide with a cleaved facet and DC bonding wires on the extended top metallization. 3D electromagnetic numerical simulations show the formation of standing waves inside the cavity due to the finite end facet reflectivities (in this case in the order of 60\% at 3 THz, very similar as for standard double metal waveguides). 
\par Light-current-voltage (LIV) characteristics, shown in Fig. \ref{fig:RF_THz}(b), display a very low threshold current density in the order of 140 A/cm\textsuperscript{2} at a heat sink temperature of 40 K, which is due to the low-loss Cu-Cu planarized waveguide and the low-dissipation superdiagonal active region \cite{forrer_photon-driven_2020}. These devices typically operate up to around 115 K in continuous wave (CW). A \SI{40}{\micro\metre} wide and 2.7 mm long ridge waveguide with cleaved facets reaches output powers in the order of  3.0 mW at 40 K and 2.4 mW at 80 K in CW at rollover. The measured powers are from a single facet and uncorrected for any absorption in the cryostat window.
\par

In recent years, the possibility to obtain frequency combs from compact, on-chip sources\cite{Chang_BowersNatPhot2022} has opened several possibilities from spectroscopy to LIDAR, remote sensing and coherent communications \cite{Marin_palomo_NatPhot2017}. 
Quantum cascade lasers in the Mid-IR and THz proved to be excellent candidates for integrated, semiconductor-based comb sources \cite{faist_quantum_2016}.  We examine now the comb properties of our lasers based on planarized waveguides. 
Free-running devices can generate frequency comb states, where the THz modes are exactly equidistantly spaced, have a fixed phase relationship and produce a single RF signal (beatnote) at the mode spacing frequency \cite{hugi_mid-infrared_2012}. A typical measurement result is shown in Fig. \ref{fig:RF_THz}(c), where the THz spectrum spans around 800 GHz and there is a single stable RF beatnote at the cavity repetition rate $f_{\mathrm{rep}}$. The measured free-running RF beatnotes in planarized waveguides can reach relatively high powers of -60 to -55 dBm at the spectrum analyzer readout, confirming improved RF properties (as described in more detail in the previous section). Free-running beatnote maps of several devices can be found in the Supplementary Material. Moreover, self-starting pure harmonic states \cite{kazakov_self-starting_2017,BurghoffOptica2020} (Turing patterns) can be observed at specific bias points, where the mode spacing is an integer multiple of the fundamental $f_{\mathrm{rep}}$ \cite{ForrerAPL2021Harmonic}. 
A typical harmonic state spectrum spanning over 1.1 THz is shown in Fig. \ref{fig:RF_THz}(d), in this case corresponding to the  third harmonic. 
\par By injecting an external RF signal to the laser cavity, it is possible to strongly modify the lasing operation both of a comb state \cite{SchneiderRF_LPR2021,hillbrand_coherent_2019} or also of a high phase noise state\cite{forrer_photon-driven_2020,gellie_injection-locking_2010}. In Fig. \ref{fig:RF_THz}(e) we show an example where injecting a strong RF signal (+32 dBm at source) close  to the natural cavity mode spacing ($f_{\mathrm{rep}}$ $\pm\leq$200 MHz) can broaden the THz emission spectrum to over 1.5 THz. These are typically not frequency comb states due to the limiting chromatic dispersion over such a wide bandwidth (as evident already from the asymmetric interferograms in standard FTIR spectrum measurements \cite{ForrerAPL2021Harmonic}), but still very useful as sources of broadband THz radiation.

\par
\begin{figure*}[h!]
\centering
\includegraphics[width=\linewidth]{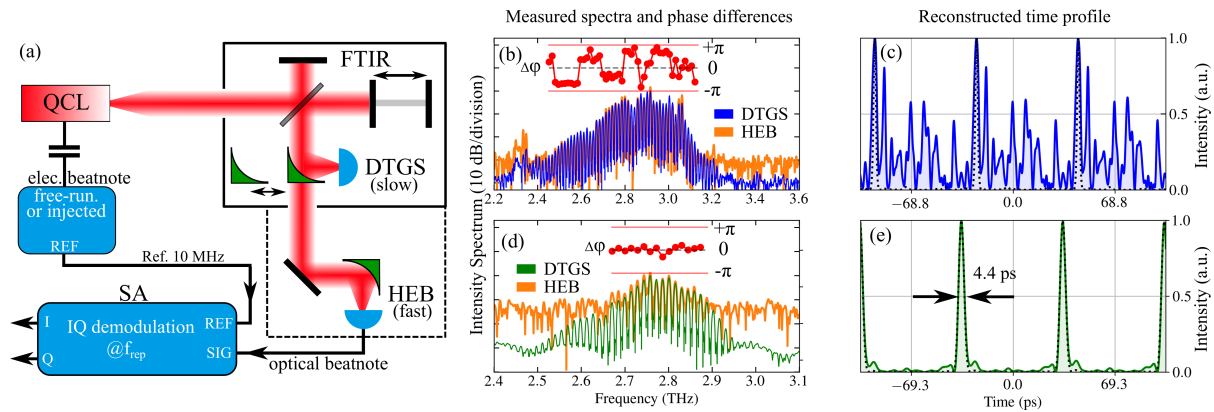}
\caption{\textbf{(a)} Schematic of the SWIFTS setup, featuring an FTIR and a hot electron bolometer (HEB) as a fast detector. \textbf{(b, c)} Measurements of a weakly-injected free-running ridge device show scattered phase differences, and the reconstructed time profile has a quasi-continuous periodic output intensity. \textbf{(d, e)} Strong RF injection (+32 dBm at the source) close to the repetition rate frequency on a ridge device close to lasing threshold results in active mode locking, producing pulses as short as 4.4 ps (close to the Fourier limit, dotted line). Here, the signal-to-noise ratio of the HEB measurement is reduced due to RF pickup problems with stronger RF injection.}
\label{fig:SWIFTS}
\end{figure*}

 In order to assess the comb coherence and retrieve the phase between adjacent modes for reconstructing the time-domain emission profile, we used Shifted Wave Interference Fourier Transform Spectroscopy (SWIFTS) \cite{burghoff_terahertz_2014,Burghoff2020}. This is a coherent beatnote technique that requires a fast detector combined with an FTIR, as illustrated in the schematic in Fig. \ref{fig:SWIFTS}(a). A hot electron bolometer (HEB) \cite{Semenov_HEB_2002} was used as the fast detector. It consists of a thin film (6 nm) NbN detecting element \cite{MartiniHEB_OPEX_2021}, with a metallic (Ti/Au) log spiral antenna for an efficient in-coupling of the incoming THz radiation. Mounting and antenna-coupling have been optimized to enhance RF performance.  During operation, it is cooled down to the superconducting state (below 10 K) and features ultrafast rise times in the order of $\sim$40 ps \cite{brundermann_terahertz_2012, Semenov_HEB_2002}. When illuminated with a THz QCL frequency comb, the optical beatnote generated between adjacent THz modes can be measured directly on the bias line of the detector and fed into a spectrum analyzer with an IQ demodulator for SWIFTS \cite{Forrer2022}.
\par Here, we present two measurement examples in two different regimes. First, we analyze the comb state of a ridge waveguide device, injection-locked to the natural $f_{\mathrm{rep}}$ = 14.540 GHz with a relatively weak reference RF signal (+5 dBm at the source) in order to stabilize the repetition. The measured spectra and relative phases are shown in Fig. \ref{fig:SWIFTS}(b), where the THz spectra measured with a slow DTGS detector (blue) and the HEB (orange) have a good overlap and comparable signal-to-noise ratio. This is attributed to the high coherence of the comb state. The relative phases are scattered across different values (with some mode groups sharing the same phase), which also depends on the amplitude of the injected RF signal. The reconstructed time profile produces a quasi-continuous periodic output with some amplitude modulation, as shown in Fig. \ref{fig:SWIFTS}(c). 
\par The second example is the same ridge waveguide device operating just above lasing threshold and driven by a strong RF injection (+32 dBm at the source) at $f_{\mathrm{inj}}$ = 14.538 GHz, close to the natural repetition rate $f_{\mathrm{rep}}$ = 14.540 GHz. The measurement results in Fig. \ref{fig:SWIFTS}(d) show that we can reach an active mode locking regime, where all the measurable modes have a flat relative phase profile, and the reconstructed time profile in Fig. \ref{fig:SWIFTS}(e) is a train of nearly Fourier-limited pulses as short as 4.4 ps. In this case, the measurement with the HEB (orange) suffers from a worse signal-to-noise ratio as compared to the DTGS (green). This is due to the limited dynamic range of the HEB (short pulses have higher peak intensities) and RF pickup problems due to the strong RF driving signal. Since only the central high-intensity modes could be measured with the HEB detector, the actual emitted THz pulses are very likely even shorter (assuming all the modes have the same phase).  

\par 
We should also emphasize that for all the frequency comb results presented in this paper (including SWIFTS), the lasing modes throughout the full measured THz emission spectrum share both the same repetition frequency $f_{\mathrm{rep}}$ and the same offset frequency $f_{\mathrm{ceo}}$, and are thus part of the same comb. In the case of matching $f_{\mathrm{rep}}$, but several different values of $f_{\mathrm{ceo}}$, the spectrum consists of several (individual) sub-combs, in which case the SWIFTS time-domain reconstruction could not have been computed for the complete emission spectrum at once. Instead, it would be sliced into the contributions of each individual sub-comb. These aspects together with a thorough description of the FM to AM transition under RF injection will be discussed elsewhere \cite{Forrer2022}.

\section*{Passive waveguide components}

\par 

\par
\begin{figure}[h!]
\centering
\includegraphics[width=0.9\linewidth]{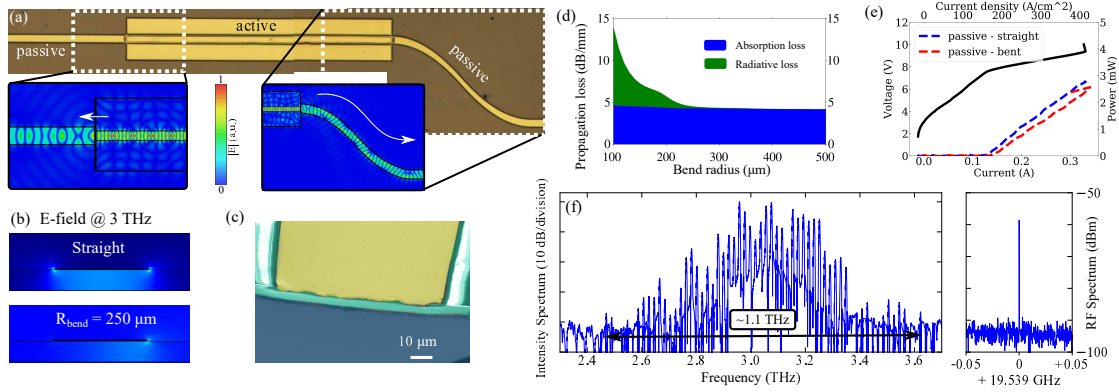}
\caption{\textbf{(a)} Optical microscope image of a ridge device co-integrated with straight and curved passive waveguide elements, consisting of a metallized stripe on top of BCB. Insets show the computed THz wave propagation with minimal scattering and bending losses at 3 THz. \textbf{(b)} Simulated electric field distribution in the passive waveguide cross-section at 3 THz. In the case of a straight waveguide, the intensity is concentrated symmetrically below the metal stripe, while it shifts to the outer side in the case of a bent path. \textbf{(c)} False color SEM images of the cleaved passive waveguide with a flat BCB end facet. \textbf{(d)} Computed passive waveguide propagation losses in dB/mm for a varying bend radius, obtained by COMSOL 2D eigenmode simulations at a frequency of 3 THz. The total losses are split into contributions from absorption loss (overlap with lossy metals and BCB) and radiative loss (bending loss), which is negligible for bend radii above \SI{230}{\micro\metre}. \textbf{(e)} The LIVs measured in pulsed (500 ns pulses, 10\% duty cycle, 20 K) show an increased threshold current density and a higher slope efficiency, both consistent with the lower end facet reflectivity. \textbf{(f)} Spectrum of a free-running device (CW, T = 20 K) in the comb regime with a bandwidth over 1 THz and a single RF beatnote above -60 dBm, indicating that the reduced cavity feedback can be beneficial for comb formation.}
\label{fig:Passive}
\end{figure}

 Another important advantage of our planarized platform is the possibility to co-integrate active and passive elements. While we have already demonstrated planar reflective and outcoupling passive antenna structures \cite{Bosco2016, Senica2020}, here we designed passive waveguides for on-chip signal routing between various elements. 
Integration of passive waveguides is critical in the development of an integrated photonic platform. Near-infrared photonic circuits are nowadays a reality, and a similar approach can be envisioned for THz frequencies. 
An optical microscope image of a fabricated device is shown in Fig. \ref{fig:Passive}(a), where an active ridge is connected to straight and bent passive waveguides on each side. The passive waveguide is a metallized stripe on top of BCB which continues beyond the active region waveguide. Full-wave 3D numerical simulations show that the optical mode is guided below the metal stripe, following also non-straight paths (see insets in Fig. \ref{fig:Passive}(a)). The reflectivity (coupling efficiency) at the active/passive waveguide interface can be tuned by the shape of the active and passive waveguides. In the simplest case of a flat active waveguide facet, a reflectivity of $R=\lvert S_{11}\rvert^2=22.5\%$ into the fundamental active waveguide mode, and a transmission of $T=\lvert S_{21}\rvert^2=57.0\%$ into the fundamental passive waveguide mode is obtained from a 3D numerical numerical simulation at a frequency of 3 THz. This gives an insertion loss of $L_\mathrm{ins}= 1-\frac{T}{1-R}=26.5\%= 1.34\ \mathrm{dB}$.
\par
We performed 2D COMSOL simulations to evaluate the passive waveguide propagation losses. For a straight waveguide (field profile in the top panel of Fig. \ref{fig:Passive}(b)), a propagation loss of around $\alpha_\mathrm{pass}=$ 9.5 cm\textsuperscript{-1} = 4.1 dB/mm at 3 THz is obtained. This originates mainly from  the non-optimized optical mode overlap with the top and bottom metal layers (a copper/gold stack), as BCB has a relatively low loss of 3 cm\textsuperscript{-1} = 1.3 dB/mm at around 3 THz \cite{bonzon2016phase}. If we consider a waveguide bend with a radius R\textsubscript{bend}, the electric field intensity is shifted towards the outer part of the passive waveguide, as shown in the bottom panel in Fig. \ref{fig:Passive}(b). Below a certain critical radius R\textsubscript{bend}, which depends on the frequency and passive waveguide width, the propagation losses start to increase exponentially due to bending losses. In Fig. \ref{fig:Passive}(c), we plot the propagation losses of a passive waveguide with a top metallic stripe width of \SI{60}{\micro\metre} at 3 THz. The radiative losses are negligible for a bend radius above around \SI{230}{\micro\metre}, slightly above two times the free-space wavelength. This value however does not impose a definitive limitation for the integration density of passive components, as it is possible to implement, for example, low-loss sharp waveguide bends with inverse-designed optimized components \cite{Liu2018}.

To characterize the passive waveguides experimentally, we performed a mechanical cleaving around 1 mm away from the active-passive waveguide interfaces on both sides and mounted the sample on a custom holder to measure the laser output from both sides of the device (straight and bent). The active region waveguide has a width of  \SI{40}{\micro\metre} and a length of 2 mm. Compared to cleaved active ridge facets, due to refractive index matching and the continued top metallized stripe, the end facet reflectivity reduces from around 60\% to around 22.5\%, as predicted by 3D numerical simulations. A false-color SEM image of the cleaved passive waveguide can be seen in Fig. \ref{fig:Passive}(c). The measured LIV curves in pulsed (500 ns pulses, 10\% duty cycle, 20 K) are shown in Fig. \ref{fig:Passive}(e). 
The peak output powers from the straight (1 mm long) and bent side (around 1.2 mm long) of the device are comparable, indicating low propagation and bending losses. The threshold current density increased to around 160 A/cm\textsuperscript{2} due to higher mirror losses. Using the mirror loss formula $\alpha_\mathrm{mirr}=-\frac{1}{2L}\ln(R_1R_2)$ and comparing a 2 mm long waveguide with cleaved facets ($R_1=R_2=60\%$) versus coupling on both sides to a passive waveguide ($R_1=R_2=22.5\%$), the mirror losses increase from 2.6 cm\textsuperscript{-1} to 7.5 cm\textsuperscript{-1}. The slope efficiency of the measured samples is increased from 11.9 mW/A to 13.6 mW/A. This is not as large as anticipated from the reduced facet reflectivity, however we need to take into account that the tested sample is coupled to a relatively long (around 1 mm) passive waveguide, which induces propagation losses. A simple back-propagation calculation gives us an upper estimate of the power emitted from the active facet (without absorption losses in the passive section): $P^{'}= P\  e^{(\alpha_\mathrm{pass}\,L)} = 7.1$ mW. Here, we used the numerical simulation result of $\alpha_\mathrm{pass}=9.5$ cm\textsuperscript{-1} and $L =$ 1 mm. This would yield an increased slope efficiency of 37.6 mW/A. Although the actual measured power is lower, it could be reached with a shorter passive waveguide, and possibly increased further by using a passive outcoupling structure as in Ref. \cite{Bosco2016}.

 
The frequency comb performance does not deteriorate, but is actually improved. In Fig. \ref{fig:Passive}(f) we showcase the largest measured comb bandwidth, broader than 1 THz (wider than any other measured sample), and a single RF beatnote stronger than -60 dBm (without using the custom RF PCB from Fig. \ref{fig:simulations}(d)).
As anticipated in the Mid-IR  by numerical results in Ref.\cite{Beiser_Schwarz_2021OptLett} and theoretical models in Ref. \cite{Humbard_burghoffOpex2022}, the product of the facet reflectivities of a QCL ridge laser affects the maximum bandwidth obtainable by a given gain medium, and a lower reflectivity should enhance the comb bandwidth. In particular, the maximum bandwidth is achieved for a product $R_1 R_2\simeq 0.2$. In our case, the lower reflectivity at the ends of the double metal cavity  due to the coupling to the passive section brings the product $R_1R_2\vert_{active}=0.36$ to $R_1R_2\vert_{act+pass}=0.062$. 


By using different combinations of cleaved and passive waveguide facets, the planarized geometry offers the possibility to change the cavity reflectivity, adding an important element to the  comb engineering toolbox. This could be expanded further by fabricating more complex dry-etched waveguide facet geometries, for example an adiabatic taper coupled to a passive waveguide, to make the reflectivity even lower. Finally, coupling broadband frequency combs to passive waveguides on the same chip, as already demonstrated in the Mid-IR\cite{Wang_passive_ACS_2022}, is an important milestone towards fully integrated active and passive THz photonic circuits and spectrometers.


\section*{Conclusions}
In conclusion, we have presented a novel platform for broadband coherent THz photonics based on high-confinement  active and passive planarized double metal waveguides. The extended top contact metallization enables bonding wires to be placed over the BCB-covered area, which results in lower waveguide losses, improved RF and thermal dissipation properties, and allows for an independent design of THz and RF waveguide properties. The fabrication of narrow waveguides acts as a fundamental mode selection mechanism and further improves heat dissipation. Free-running broadband frequency combs over 800 GHz and harmonic states over 1.1 THz are observed. Driven with an additional external RF signal, broadband THz emission spectra over nearly 1.6 THz, and actively mode-locked pulses as short as 4.4 ps can be generated on demand. The co-integration of passive elements with custom on-chip guiding, reflection and outcoupling properties complete the diverse integrated THz photonics toolbox enabled by our planarized platform. 

\section*{Methods}\label{method}

\subsection{Planarized waveguide fabrication}
From MBE-grown wafers, samples with a typical size of (9x10) mm\textsuperscript{2} were cleaved. A metallic stack of Ta(5nm)/Cu(250nm)/Ti(50nm)/Au(500nm) was deposited on the sample and on a carrier n+ GaAs substrate by electron beam evaporation. These were bonded using thermocompression wafer bonding at a temperature of 320°C and a pressure of 5 MPa for 15 minutes in vacuum. After a mechanical thinning and several wet etch steps to expose the active region, the waveguides were dry-etched in a Cl$_2$/H$_2$ inductively-coupled plasma (ICP) using SiNx as a hardmask. Subsequently, AP3000 was spun on the sample as an adhesion layer for BCB (3000 RPM for 30 s). Cyclotene 3022-57 (BCB) was both spin coated (5000 RPM for 50 s) and baked sequentially five times (4x softbake at 210°C and 1x hardbake at 250°C for 2h). This resulted in a total BCB thickness of around \SI{25}{\micro\metre} to ensure a smooth profile across the sample. Using Reactive Ion Etching (RIE), the BCB was etched to match the height of the active region waveguide (\SI{10.4}{\micro\metre}). Finally, the extended top metallization (same metallic stack as the wafer-bonded side) with a typical width of \SI{300}{\micro\metre} was defined with photolithography and lift-off, spanning over the active waveguides and the BCB-covered area on the sides. The passive waveguides were fabricated over BCB-covered areas in this final step as well.

\subsection{RF mounting}
The custom two-layer RF PCB was fabricated on a Rogers 4350 substrate with a thickness of 0.8 mm. The coplanar waveguide has a width of \SI{800}{\micro\metre} and is separated from the side ground planes by \SI{100}{\micro\metre}. A series of vias connecting to the ground plane improves the RF performance and the isolation between RF and DC ports. The PCB is screwed to a copper laser submount and mounted on a cryostat cold finger. It is connected with a standard RF connector (HK-LR-SR2(12)) to low-loss semi-rigid RF cables (SUCOFORM 86 FEP), accessible from the outside of the cryostat.

\begin{addendum}

 \item[Acknowledgements] The authors gratefully acknowledge funding from the ERC Grant CHIC (No. 724344) and in part by the NATO Science for Peace and Security Programme under grant G5721 THESEUs. We also thank  Tabea Bühler, Sebastian Gloor and Johannes Hillbrand for technical help.
 \item[Competing Interests] The authors declare that they have no competing financial interests.
 \item[Authors contributions] U.S. and G.S. conceived the idea. U.S. designed and fabricated the devices, carried out all the measurements, analysed experimental data and performed numerical simulations under the supervision of G.S. and J.F.. A. F. designed the RF PCB board and built the SWIFTS setup. U.S. and T.O. developed the planarized waveguide fabrication process. P.M. performed some of the device characterizations. S.C. and G.T. provided the HEB detectors, A.F. optimized the HEB RF coupling. M.B. performed the epitaxial growth. U.S. and G.S. wrote the manuscript. All authors discussed the results and commented on the manuscript.
 
 \item[Correspondence]  *Correspondence should be addressed to U. Senica (email: usenica@phys.ethz.ch) and G. Scalari (email: scalari@phys.ethz.ch).
\end{addendum}

\section*{References}\label{References}
\bibliography{GS_bib_PlanarizedWG.bib,GS_bib.bib}





\end{document}